\documentclass[a4paper]{article}
\usepackage{ASVspoof2021}
\usepackage[ruled,vlined,algo2e]{algorithm2e}
\usepackage[bottom]{footmisc}
\usepackage{epsfig,amssymb,amsmath,rotating}
\ninept
\usepackage{algpseudocode} 
\usepackage{pgfplots,balance}
\usepackage{grffile}
\usepackage{float}
\usepackage{lipsum}
\usepackage{times}
\usepackage{latexsym}
\usepackage{amssymb}
\usepackage{url}
\usepackage{graphicx}
\usepackage{epstopdf}
\usepackage{mwe}
\usepackage[export]{adjustbox}
\usepackage{tikz}
\usepackage{algorithm}
\usepackage{kantlipsum}
\usepackage{xpatch,xcolor}
\usepackage{graphicx}
\usepackage{afterpage}
\usepackage{balance}
\usepackage{cite}
\usepackage{xcolor}
\makeatletter
\usepackage{tabularx}
\usepackage{multirow}
\usepackage{hyperref}
\usepackage{arydshln}
\hypersetup{
    colorlinks=true,
    linkcolor=purple,
    filecolor=magenta,      
    urlcolor=purple,
}

\makeatother

\title{End-to-End Spectro-Temporal Graph Attention Networks for \\  Speaker Verification Anti-Spoofing and Speech Deepfake Detection}

\makeatletter
\def\name#1{\gdef\@name{#1\\}}
\makeatother
\name{
  {\em Hemlata Tak$^1$, Jee-weon Jung$^2$, Jose Patino$^1$, Madhu Kamble$^1$,}\\ 
  {\em Massimiliano Todisco$^1$ and Nicholas Evans$^1$}
}
\address{$^1$EURECOM, Sophia Antipolis, France, $^2$Naver Corporation, South Korea \\
{\small \tt lastname@eurecom.fr, jeeweon.jung@navercorp.com} }

\begin{document}
\maketitle
\begin{abstract}
Artefacts that serve to distinguish bona fide speech from spoofed or deepfake speech are known to reside in specific sub-bands and temporal segments. Various approaches can be used to capture and model such artefacts, however, none works well across a spectrum of diverse spoofing attacks. Reliable detection then often depends upon the fusion of multiple detection systems, each tuned to detect different forms of attack. In this paper we show that better performance can be achieved when the fusion is performed within the model itself and when the representation is learned automatically from raw waveform inputs. The principal contribution is a spectro-temporal graph attention network (GAT) which learns the relationship between cues spanning different sub-bands and temporal intervals. Using a model-level graph fusion of spectral (S) and temporal (T) sub-graphs and a graph pooling strategy to improve discrimination, the proposed RawGAT-ST model achieves an equal error rate of 1.06\% for the ASVspoof 2019 logical access database. This is one of the best results reported to date and is reproducible using an open source implementation.
\end{abstract}

\section{Introduction}
\label{sec:intro}

It is well known that cues indicative of spoofing attacks reside in specific sub-bands or temporal segments~\cite{sahidullah2015comparison,sriskandaraja2016investigation,witkowski2017audio,nagarsheth2017replay,lin2018replay,chettri2019ensemble,yang2019significance,garg2019subband,tomi2020subband,odyssey2020CQCC,IS2020LFCC,tak2021rawnet,tak2021graph}.   
Prior work~\cite{odyssey2020CQCC,tomi2020subband,chen2020generalization,IS2020LFCC,tak2021rawnet} showed that these can be learned automatically using a model with either spectral or temporal attention.
Our most recent work~\cite{tak2021graph} showed the merit of graph attention networks (GATs) to learn the relationships between cues located in different sub-bands or different temporal intervals.
We observed that different attention mechanisms spanning either the spectral or the temporal domain, work more or less well for different sets of spoofing attack and that the benefit of both can be exploited through their fusion at the score level.
We also found that neither model works as well on its own for the full set of diverse spoofing attacks in the ASVspoof 2019 logical access database.
Motivated by psychoacoustics studies~\cite{Youngberg78,Brown91,schorkhuber2014matlab,todisco2016new} which show the power of the human auditory system to select simultaneously the most discriminative spectral bands and temporal segments for a given task, and inspired by the power of GATs to model complex relationships embedded within graph representations, we have explored fusion within the model itself (earlier fusion).
The idea is to extend the modelling of relationships between \emph{either} different sub-bands \emph{or} different temporal segments using separate models and attention mechanisms to the modelling of relationships spanning different spectro-temporal intervals using a GAT with combined spectro-temporal attention (GAT-ST).
The approach facilitates the aggregation of complementary, discriminative information simultaneously in both domains. 

In a further extension to our past work, the GAT-ST reported in this paper operates directly upon the raw waveform.
Such a fully end-to-end approach is designed to maximise the potential of capturing discriminative cues in both spectral and temporal domains. 
Inspired by work in speaker verification~\cite{jung2018avoiding, ravanelli2018speaker,kamble2020advances,jung2020improved} and in building upon our end-to-end anti-spoofing solution reported in~\cite{tak2021rawnet}, the proposed RawGAT-ST model uses a one-dimensional sinc convolution layer to ingest raw audio. 
The principal contributions of this paper hence include:
\begin{itemize}

\item a fully end-to-end architecture comprising feature representation learning and a GAT;

\item a novel spectro-temporal GAT which learns the relationships between cues at different sub-band and temporal intervals;

\item a new graph pooling strategy to reduce the graph dimension and to improve discrimination;
\item an exploration of different model-level, graph fusion strategies.
\end{itemize}

The remainder of this paper is organized as follows. Section~\ref{section:Related work} describes related works.  Section~\ref{section:GAT_spoofing} provides an introduction to GATs and describes how they can be applied to anti-spoofing.   The proposed RawGAT-ST model is described in Section~\ref{section:End-to-end GAT}. Experiments and results are presented in Sections~\ref{section:exp_setup} and~\ref{sec:experimental results}.  Finally, the paper is concluded in Section~\ref{sec:conclusion}.


\section{Related works}
\label{section:Related work}
In recent years, graph neural networks (GNNs)~\cite{gori2005new,scarselli2008graph,bronstein2017geometric,hamilton2017inductive,wu2020comprehensive} have attracted growing attention, especially variants such as graph convolution networks (GCNs)~\cite{Kipf2017Semi-supervisedNetworks} or GATs~\cite{velivckovic2017graph}. 
A number of studies have shown the utility and appeal of graph modeling~\cite{zhang2019few,liu2020graphspeech,jung2020graph,tzirakis2021multi} for various speech processing tasks.
Zhang et al.~\cite{zhang2019few} applied GCNs to a few-shot audio classification task to derive an attention vector which helps to improve the discrimination between different audio examples.  Jung et al.~\cite{jung2020graph} demonstrated the use of GATs as a back-end classifier to model the relationships between enrollment and test utterances for a speaker verification task. 
Panagiotis et al.~\cite{tzirakis2021multi} used GCNs to exploit the spatial correlations between the different channels (nodes) for a multi-channel speech enhancement problem.

Our previous work~\cite{tak2021graph} demonstrated how GATs can be used to model spoofing attack artefacts.
This is achieved using a self attention mechanism to emphasize the most informative sub-bands or temporal intervals and the relationships between them.
We applied GATs separately to model the relationships in either spectral (GAT-S) or temporal (GAT-T) domains 
and demonstrated their complementarity through a score-level fusion (i.e., \emph{late} fusion).
Our hypothesis is that the integration of these two models (i.e., \emph{early} fusion) has better potential to leverage complementary information (spectral and temporal) and further boost performance while using a \emph{single} model.
We present a summary of our original GAT-S/T approach in the next section. 
The new contribution is described in Section~4. 



\begin{figure}[!t]
  \centering
  \includegraphics[width=\linewidth]{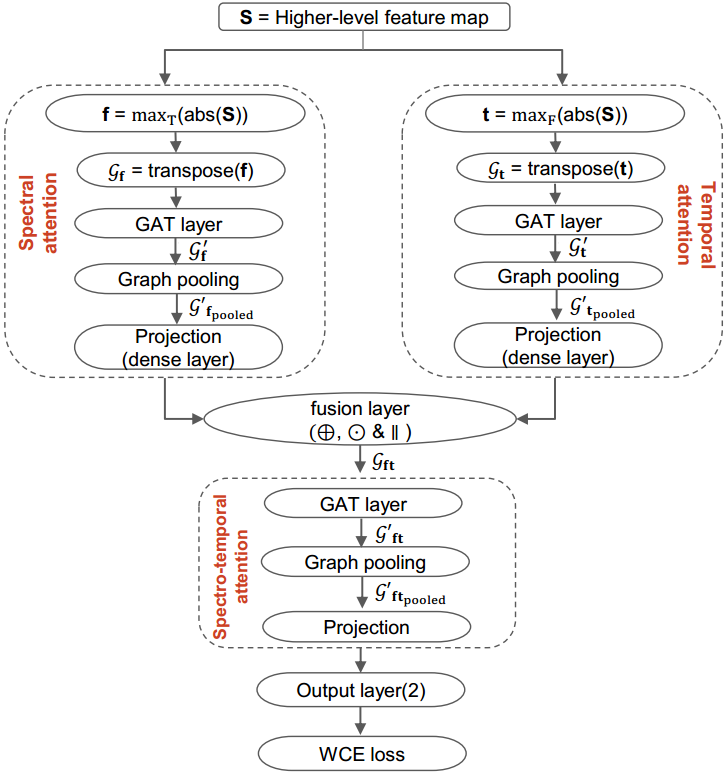}
  \vspace{-0.55cm}
  \caption{
    The proposed RawGAT-ST model architecture. The top-left spectral attention block captures  discriminative spectral information. The top-right temporal attention block captures  discriminative temporal information. Fusion is performed at the model level (middle ellipse). The bottom block comprises the spectro-temporal graph attention model.}
   \label{fig:GAT pipeline}
\end{figure}

\begin{figure*}[!ht]
  \centering
  \includegraphics[width=17cm]{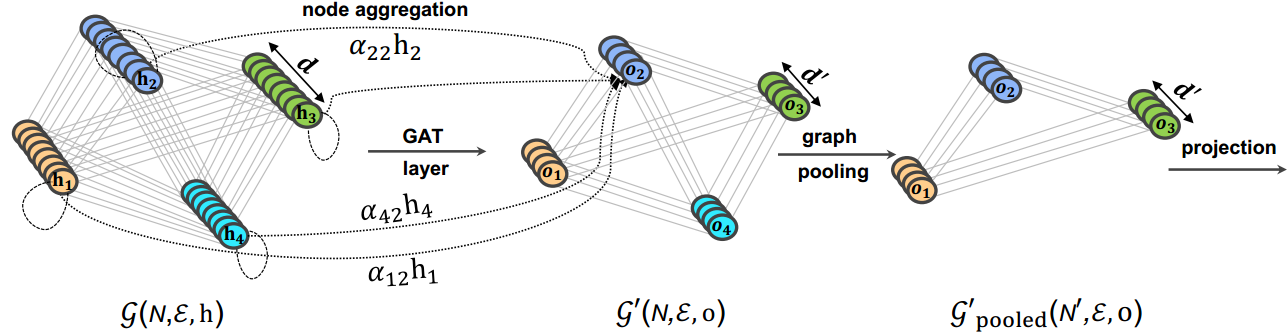}

  \caption{An illustration of the proposed graph-based process including a graph attention layer and a graph pooling layer for an input graph with 4 nodes. 
  $\textrm{h}_{n}\in\mathbb{R}^{d}$ represents the node representation $n \in N$.
  The graph attention layer estimates attention weights $\alpha$ between each node pair, aggregating node information, and projecting to a new representation space $\textrm{o}_{n}\in\mathbb{R}^{d'}$ with lower output dimensionality $d'$.
  The graph pooling layer reduces the number of nodes to improve discrimination. 
  }
   \label{fig:GAT process dia}
 
\end{figure*}





	
	

\section{Graph attention networks for anti-spoofing}
\label{section:GAT_spoofing}
The work in~\cite{tak2021graph} demonstrated the use of GATs to model the relationships between artefacts in different sub-bands or temporal intervals. 
The GAT learning process is illustrated in Algorithm~\ref{alg: GAT process} and is described in the following.
A graph is defined as:
\begin{equation}
 \mathcal{G}(N,\mathcal{E},\textrm{\textbf{h}}),
\end{equation}
where $N$ is the set of nodes, $\mathcal{E} $ represents the edges between all possible node connections, including self-connections.
$\mathcal{G}$ is formed from a higher-level feature representation $\textbf{\textrm{h}}$ (e.g.\ the output feature map of a residual network), where $\textbf{\textrm{h}}\in\mathbb{R}^{N \times d}$.  The number of nodes in the set $N$ is the number of spectral sub-bands remaining after temporal averaging applied to each channel or, conversely, the number of temporal frames after spectral averaging.
The feature dimensionality of each node $d$ is equal to the number of channels in the feature map.  

The GAT layer operates upon an input graph $\mathcal{G}$ to produce an output graph $\mathcal{G'}$. 
It aggregates neighboring nodes using learnable weights via a self-attention mechanism. 
Node features are projected onto another representation learned from the minimisation of a training loss function using an affine transform (dense layer). 
Nodes are aggregated with attention weights which reflect the connective strength (relationship) between a given node pair.




The output graph $\mathcal{G'}$ comprises a set of nodes $\textrm{o}$, where each node $\textrm{o}_n$ is derived according to:
\begin{equation}
    \textrm{o}_n=\textrm{SeLU}(\textrm{BN}(W_{\textrm{att}}(m_{n}) + W_{\textrm{res}}({{\textrm{h}}}_{n}))),
    \label{Eq.: node propagation}
\end{equation}
where SeLU refers to a scaled exponential linear unit~\cite{klambauer2017self} activation function, BN refers to batch normalisation~\cite{Ioffe2015BatchShift}, ${m}_{n}$ is the aggregated information for node $n$, and $ \textrm{h}_{n}\in\mathbb{R}^{d}$ represents the feature vector of node $n\in N$.  Each output node $\textrm{o}_n \in\mathbb{R}^{d'}$ has a target dimensionality $d'<d$.
$W_{\textrm{att}}$ projects the aggregated information for each node $n$ to the target dimensionality~$d'$. $W_{\textrm{res}}$ projects the residual (skip) connection output to the same dimension.

The information from neighboring nodes is aggregated via self-attention according to:
\begin{equation}
    m_n=\sum_{u \in \mathcal{M}(n) \cup \{ n \} }\alpha_{u,n}{\textrm{h}}_{u},
    \label{Eq:node aggregation}
\end{equation}
where $\mathcal{M}(n)$ refers to the set of neighbouring nodes for node $n$, and $\alpha_{u,n}$ refers to the attention weight between nodes $u$ and~$n$.
The GAT layer assigns learnable self-attention weights $\alpha$ to each edge. 
Weights reflect how informative one node is of another, where higher weights imply a higher connective strength.

The attention weights are derived according to:
\begin{equation}
    \alpha_{u,n}=\frac{{\operatorname{exp}}(W_{\textrm{map}}({\textrm{h}}_{n} \odot {\textrm{h}}_{u}))}
    {\sum_{w \in \mathcal{M}(n) \cup \{ n \}}  {\operatorname{exp}}(W_{\textrm{map}}({\textrm{h}}_{n} \odot {\textrm{h}}_{w}))  },
\end{equation}
where $W_{\textrm{map}}\in\mathbb{R}^{d}$ is the learnable map applied to the dot product and where $\odot$ denotes element-wise multiplication. 

\begin{algorithm}[t]
\SetAlgoLined
 \textbf{Input}: \(\mathcal{G}\) (\textit{ N}, \(\mathcal{E}\), \textrm{\textbf{h}})\\
 
 $\textrm{h}_{n}$ $\in$ $\mathbb{R}^{d}$, $n \in N$  \\
    
    \For {$n$ $\in$ $N$} 
            {
         
            $\forall u \in \mathcal{M}(n)\cup \{n\}$, $\mathcal{M}(n)$ \textrm{(set of nodes neighboring n )}\\
            
            $\alpha_{u,n}$ $\leftarrow$ \textrm{softmax}$(W_{\textrm{map}}(\textrm{h}_{n} \odot \textrm{h}_{u}))$\\

            $m_{n}$ $\leftarrow$ $\sum_{u}{\alpha_{u,n}\textrm{h}_{u}}$, \textrm{node aggregation  w.r.t Eq.~\ref{Eq:node aggregation}}\\
            
            $\textrm{o}_{n}$ $\leftarrow$ \textrm{SeLU}$(\textrm{BN}(W_{\textrm{att}}(m_n) + W_{\textrm{res}}(\textrm{h}_{n})))$, $\textrm{o}_{n}$ $\in$ $\mathbb{R}^{d'}$\\

         }
\textbf{Output}: $\mathcal{G'}$ $\leftarrow$ $\textrm{o}$\\
\caption{The GAT learning process}
\label{alg: GAT process}
\end{algorithm}

 
 
    
         
         
         
        

\section{RawGAT-ST model for anti-spoofing and speech deepfake detection}
\label{section:End-to-end GAT}




In this section, we introduce the proposed raw GAT with spectro-temporal attention (RawGAT-ST) model. 
It comprises four stages: 
i)~learning higher-level semantic feature representations in truly end-to-end fashion by operating on the raw waveform; 
ii) a novel graph attention module with spectro-temporal attention;
iii) a new graph pooling layer for discriminative node selection;
iv) model-level fusion.
The architecture is illustrated in Fig.~\ref{fig:GAT pipeline} whereas
a summary of the network configuration is shown in Table~\ref{Tab:RawGAT details}.

\subsection{Front-end (higher-level) feature representation}

In contrast to our prior work~\cite{tak2021graph} which used hand-crafted features, the RawGAT-ST model operates directly upon the raw waveform~\cite{tak2021rawnet,hua2021towards}. 
The literature shows that solutions based upon a bank of sinc filters are particularly effective in terms of both convergence stability and performance~\cite{ravanelli2018speaker,jung2019rawnet,tak2021rawnet}. 
Accordingly, we use a sinc convolution layer for front-end feature learning similar to that reported in our previous work~\cite{tak2021rawnet}.
It performs time-domain convolution of the raw waveform with a set of parameterized sinc functions which correspond to rectangular band-pass filters~\cite{quatieri2006discrete,deller1993discrete}. 
The centre frequencies of each filter in the filterbank are set according to a mel-scale.
For all work reported in this paper, rather than learning cut-in and cut-off frequencies of each sinc filter, we used fixed cut-in and cut-off frequencies to alleviate over-fitting to training data.
 
The output of the sinc layer is transformed to a time-frequency representation by adding one additional channel dimension. 
The result is fed to a 2-dimensional (2D) residual network~\cite{he2016deep} to learn higher-level feature representations $\textbf{S}\in\mathbb{R}^{C \times F\times T}$ where $C$, $F$ and $T$ refers to the number of channels, frequency bins and time samples respectively.
We use a residual neural network identical to that reported in~\cite{tak2021rawnet}, except for the use of a 2D CNN instead of 1D CNN.
A summary of the configuration is shown in Table~\ref{Tab:RawGAT details}. 
Each residual block layer consists of batch normalization (BN) with SeLU activation units, 2D convolution and a final max-pooling layer for data downsampling.  

\begin{table}[!ht]
\normalsize
	\centering

	\caption{The details of RawGAT-ST model architecture. Numbers denoted in Sinc layer \& Conv layer refers to (filter size, stride, number of filters).
	The output size are refers to (CNN channels, Freq, Time). Separate GAT layers are use for spectral and temporal attention blocks.}
	\renewcommand{\arraystretch}{1.4}
	\vspace{0.4cm}
	\resizebox{\linewidth}{!}{
	\begin{tabular}{p{55pt}p{55pt}p{55pt}p{55pt}}
	\hline
	\multicolumn{1}{c}{Layer} & \multicolumn{2}{c}{Input: 64600 samples} & \multicolumn{1}{c}{Output shape}\\
	\hline\hline
	\multicolumn{1}{c}{Sinc layer} & \multicolumn{2}{c}{Conv-1D(129,1,70)} & \multicolumn{1}{c}{(70,64472)}\\
	
	\multicolumn{1}{c}{} & \multicolumn{2}{c}{add channel (TF representation)} & \multicolumn{1}{c}{(1,70,64472)}\\
	
	\multicolumn{1}{c}{} & \multicolumn{2}{c}{Maxpool-2D(3)} & \multicolumn{1}{c}{(1,23,21490)}\\
	
	\multicolumn{1}{c}{} & \multicolumn{2}{c}{BN \& SeLU} & \multicolumn{1}{c}{}\\
	\hline
	
	\multicolumn{1}{c}{Res block} & 
        {$ \left \{ 
	    \begin{array}{c}
	   
	    \text{Conv-2D((2,3),1,32)}\\
	    \text{BN \& SeLU }\\
	    \text{Conv-2D((2,3),1,32)}\\
	    \text{Maxpool-2D((1,3))}\\
	    \end{array} \right \} 
	    \times $}
	    &\hspace{1.55cm}2
	& \multicolumn{1}{c}{(32,23,2387)}\\
	\hline
	
	\multicolumn{1}{c}{Res block} & 
        {$ \left \{ 
	    \begin{array}{c}
	   
	    \text{Conv-2D((2,3),1,64)}\\
	    \text{BN \& SeLU }\\
	    \text{Conv-2D((2,3),1,64)}\\
	    \text{Maxpool-2D((1,3))}\\
	    \end{array} \right \} 
	    \times $}
	    &\hspace{1.55cm}4
	& \multicolumn{1}{c}{(64,23,29)}\\
	\hline

 \multicolumn{2}{c}{Spectral-attention}& \multicolumn{2}{c}{Temporal-attention}\\ \cline{1-4}
\multicolumn{2}{c}{$\textrm{max}_{\textrm{T}}(\textrm{abs}())=(64,23)$} & \multicolumn{2}{c}{$\textrm{max}_{\textrm{F}}(\textrm{abs}())=(64,29)$}\\

 \multicolumn{2}{c}{$\textrm{GAT layer =(32,23)}$} & \multicolumn{2}{c}{$\textrm{GAT layer =(32,29)}$}\\
 	
\multicolumn{2}{c}{$\textrm{Graph pooling=(32,14)}$} & \multicolumn{2}{c}{$\textrm{Graph pooling=(32,23)}$}\\
  	
\multicolumn{2}{c}{$\textrm{Projection=(32,12)}$}  & \multicolumn{2}{c}{$\textrm{Projection=(32,12)}$}\\
 \hline
 
 \multicolumn{1}{c}{} & \multicolumn{2}{c}{element-wise addition} & \multicolumn{1}{c}{(32,12)}\\
 
 \multicolumn{1}{c}{Fusion layer} & \multicolumn{2}{c}{element-wise multiplication} & \multicolumn{1}{c}{(32,12)}\\
 
 \multicolumn{1}{c}{} & \multicolumn{2}{c}{concatenation (along feature dim)} & \multicolumn{1}{c}{(64,12)}\\
	\hline
	
\multicolumn{1}{c}{Spectro-} & \multicolumn{2}{c}{GAT layer} & \multicolumn{1}{c}{(16,12)}\\

\multicolumn{1}{c}{temporal} & \multicolumn{2}{c}{Graph pooling} & \multicolumn{1}{c}{(16,7)}\\

\multicolumn{1}{c}{attention} & \multicolumn{2}{c}{Projection (along feature dim)} & \multicolumn{1}{c}{(1,7)}\\
	\hline	
\multicolumn{1}{c}{Output} & \multicolumn{2}{c}{FC(2)} & \multicolumn{1}{c}{2}\\	
	
	\hline
	\end{tabular}
	}
	\label{Tab:RawGAT details}
	\vspace{-0.45cm}
\end{table}

\begin{figure*}[!h]
 \centering
 \includegraphics[width=16cm]{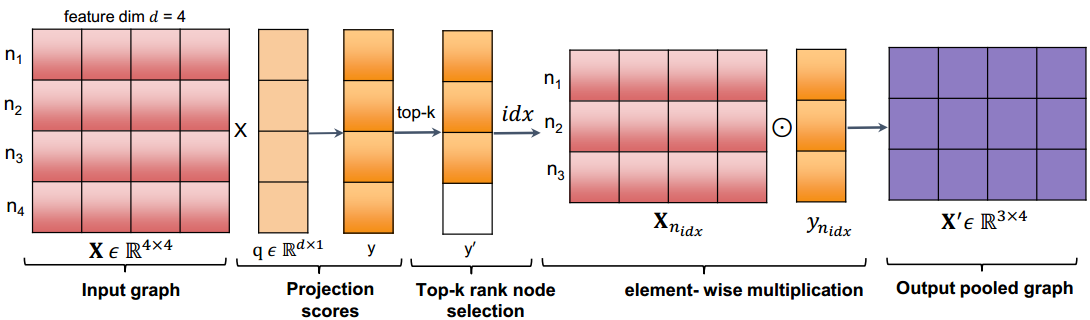}
  \caption{
    An illustration of the graph pooling layer~\cite{gao2019graph}. The input graph $\textbf{X}\in\mathbb{R}^{N \times d}$ has 
    $4$ nodes and feature dimensionality $d=4$. 
    $q\in\mathbb{R}^{d \times 1}$ is a learnable projection vector. Projection scores $y$ are obtained from the
    dot product between $q$ and the input feature vector $X_{n}$ for each node. The indices $idx$ corresponding to the nodes with the top-k highest projection scores are used to form a new pooled graph $\textrm{\textbf{X}}^\prime$ from the element-wise multiplication of ${\textbf{\textrm{X}}_{n_{idx}}}$ with $\textrm{sigmoid}({y_{n_{idx}}})$ (Eq.~\ref{Eq:sub graph}).
}
\label{Fig:Gpooling layer}
\end{figure*}

\subsection{Spectro-temporal attention}

The new approach to bring spectro-temporal graph attention to a single model is a core contribution of this work.
An overview of the proposed RawGAT-ST model architecture is 
illustrated in Fig.~\ref{fig:GAT pipeline}. 
The input to the RawGAT-ST is a higher-level feature map (top of Fig.~\ref{fig:GAT pipeline}).  The RawGAT-ST model itself comprises three principal blocks, each of which contains a single GAT layer: a spectral attention block (top-left of Fig.~\ref{fig:GAT pipeline}); a temporal attention block (top-right); a final spectro-temporal attention block (bottom).
The spectral and temporal attention blocks have the goal of identifying spectral and temporal cues.  The third, spectro-temporal attention block operates upon the pair of resulting graphs to model the relationships spanning both domains.  
All three blocks contain GAT and graph pooling layers. 
The operation of both processes is illustrated in Fig.~\ref{fig:GAT process dia}. The GAT layer operates upon an input graph $\mathcal{G}(N,\mathcal{E}, \textrm{h})$ and produces an output graph $\mathcal{G'}(N,\mathcal{E}, \textrm{o})$ by aggregating node information using self-attention weights between node pairs.  
The example in Fig.~\ref{fig:GAT process dia} shows the operation upon a 4-node input graph where the GAT layer reduces the node dimension from $d$ to $d'$.




This process is first applied separately with attention to either spectral or temporal domains to model the relationships between spectral or temporal artefacts present in the different sub-band or temporal intervals.  
Both spectral and temporal blocks operate upon 
the higher-level feature map $\textbf{S}\in\mathbb{R}^{C \times F\times T}$.
%
The spectral and temporal attention blocks first collapse temporal and spectral information respectively to a single dimension via max-pooling before the GAT layer. 
For the spectral attention block, max-pooling is applied to absolute values across the temporal dimension thereby giving a spectral feature map 
$\textrm{\textbf{f}}\in\mathbb{R}^{C \times F}$:
\begin{equation}
    \textbf{f}=\max_{\textrm{T}}(\textrm{abs}(\textbf{S})),
    \label{Eq: spectral}
    \vspace{0.2cm}
\end{equation}
The temporal attention block operates instead across the spectral dimension giving a temporal feature map 
$\textrm{\textbf{t}}\in\mathbb{R}^{C \times T}$:
\begin{equation}
   \textbf{t}=\max_{\textrm{F}}(\textrm{abs}(\textbf{S})),
   \label{Eq: temporal}
\end{equation}

\noindent Since $\textrm{\textbf{S}}$ is derived from temporal data and hence contains both positive and negative values, use of absolute values in Eqs.~\ref{Eq: spectral} and~\ref{Eq: temporal} prevents meaningful negative-valued data from being discarded.

Graphs 
$\mathcal{G}_{\textrm{\textbf{f}}}\in\mathbb{R}^{N_\textbf{f} \times d}$
and 
$\mathcal{G}_{\textrm{\textbf{t}}}\in\mathbb{R}^{N_\textbf{t} \times d}$
are then constructed from the transpose of $\textbf{f}$ and $\textbf{t}$ feature maps, respectively.  
$\mathcal{G}_{\textrm{\textbf{f}}}$ contains a set of $23$ nodes (the number of spectral bins) whereas $\mathcal{G}_{\textrm{\textbf{t}}}$ contains a set of $29$ nodes (the number of temporal segments).  Both graphs have a common dimensionality of $d=64$.
%
Separate GAT layers are then applied to both
$\mathcal{G}_{\textrm{\textbf{f}}}$ and $\mathcal{G}_{\textrm{\textbf{t}}}$  
to model the relationships between different sub-bands and temporal segments 
thereby producing a pair of new output graphs 
$\mathcal{G}_{\textrm{\textbf{f}}}^{'}\in\mathbb{R}^{N_{\textbf{f}} \times d'}$ and $\mathcal{G}_{\textrm{\textbf{t}}}^{'}\in\mathbb{R}^{N_{\textbf{t}} \times d'}$
with a common, reduced output feature dimensionality $d'=32$. 

\subsection{Graph pooling}
\label{section:Gpool}
As illustrated in Fig.~\ref{fig:GAT pipeline}, a graph pooling layer is included in all three GAT blocks.  
Graph pooling generates more discriminative graphs by selecting a subset of the most informative nodes. 
In the example illustrated in Fig.~\ref{fig:GAT process dia} the graph pooling layer reduces the number of nodes from 4 to~3.

The approach is based upon the graph pooling layer in the recently proposed Graph U-Net architecture~\cite{gao2019graph}.  An example is illustrated in Fig.~\ref{Fig:Gpooling layer}.  The process is described in the following.
Graph pooling uses a learnable projection vector $q\in\mathbb{R}^{d\times1}$.  The dot-product between the input feature vector for each node $X_{n}\in\mathbb{R}^{1\times d}$ where, $n \in N$, and $q$ gives projection scores $y_{n}$:

\begin{equation}
    y_{n}=X_{n}\cdot q,
    \label{eq: proj}
    \vspace{0.1cm}
\end{equation}

Nodes in the input graph $\textbf{X}$ corresponding to the top-$k$-hot vector $y'$ are then retained according to element-wise multiplication: 
\vspace{-0.1cm}
\begin{equation}
    \textrm{\textbf{X}}^{\prime}={\textrm{X}_{n_{idx}}} \odot \textrm{sigmoid}({y_{n_{idx}}}),
    \label{Eq:sub graph}
\end{equation}

\noindent where the pooling ratio $k$ is a hyperparameter, and where $\textrm{X}_{n_{idx}}$ and ${y_{n_{idx}}}$ are the node features and projection scores selected corresponding to the highest top-$k$ indices, $idx$. All other nodes are purged.
The complete process is illustrated in Fig.~\ref{Fig:Gpooling layer} for an input graph $\textrm{\textbf{X}}$ with $4$ nodes, each of dimension $d=4$.
Finally, pooled graph representations $\mathcal{G'}_{{\textrm{\textbf{f}}}_{\textrm{pooled}}}$ and $\mathcal{G'}_{{\textrm{\textbf{t}}}_{\textrm{pooled}}}$ are generated from the original spectral $\mathcal{G'}_{\textrm{\textbf{f}}}$ and temporal $\mathcal{G'}_{\textrm{\textbf{t}}}$ output graphs, respectively. 
Since spectral and temporal pooled graphs $\mathcal{G'}_{{\textrm{\textbf{f}}}_{\textrm{pooled}}}$ and $\mathcal{G'}_{{\textrm{\textbf{t}}}_{\textrm{pooled}}}$ have different number of nodes, $14$ for $\mathcal{G'}_{{\textrm{\textbf{f}}}_{\textrm{pooled}}}$ and $23$ for $\mathcal{G'}_{{\textrm{\textbf{t}}}_{\textrm{pooled}}}$. Hence, both graph nodes are projected into the same dimensional space using an affine-transform to match the input node dimensionality of the fusion layer.

\subsection{Model-level fusion}
\label{sec:fusion}


Model-level fusion (ellipse in Fig.~\ref{fig:GAT pipeline}) is used to exploit complementary information captured by the spectral and temporal attention graphs. 
We studied three different approaches to model-level fusion :
\vspace{-0.3cm}
\begin{equation}
\mathcal{G}_{{\textbf{ft}}}=\left\{
\begin{aligned}
\mathcal{G'}_{{\textrm{\textbf{f}}}_{\textrm{pooled}}}  &  \oplus & \mathcal{G'}_{{\textrm{\textbf{t}}}_{\textrm{pooled}}} \\
\mathcal{G'}_{{\textrm{\textbf{f}}}_{\textrm{pooled}}} &  \odot & \mathcal{G'}_{{\textrm{\textbf{t}}}_{\textrm{pooled}}} \\
\mathcal{G'}_{{\textrm{\textbf{f}}}_{\textrm{pooled}}} & \ ||  &  \mathcal{G'}_{{\textrm{\textbf{t}}}_{\textrm{pooled}}}
\end{aligned}\right.
\label{Eq: Gst}
\end{equation}


\noindent where the fused graph $\mathcal{G}_{\textbf{ft}}\in\mathbb{R}^{{N_{\textbf{ft}}} \times {d_{\textbf{ft}}}}$ is generated from one of the fusion approaches in ~Eq.~\ref{Eq: Gst}.  It acts to combine the spectral pooled graph $\mathcal{G'}_{{\textrm{\textbf{f}}}_{\textrm{pooled}}}$ with the temporal pooled graph $\mathcal{G'}_{{\textrm{\textbf{t}}}_{\textrm{pooled}}}$. A fused graph $\mathcal{G}_{\textbf{ft}}$ contains a set of $12$ nodes and feature dimensionality $d_{\textbf{ft}}=32$.
The three different operators in Eq.~\ref{Eq: Gst} are element-wise addition~$\oplus$, 
multiplication~$\odot$ and concatenation~$||$. 
A third GAT layer is then applied to $\mathcal{G}_{{\textbf{ft}}}$ to produce output graph $\mathcal{G'}_{\textbf{ft}}\in\mathbb{R}^{N_{\textbf{ft}} \times d'_{\textbf{ft}}}$, where $d'_{\textbf{ft}}=16$ is the output feature dimensionality.
Graph pooling is then applied one last time to generate a pooled graph $\mathcal{G'}_{\textbf{ft}_{\textrm{pooled}}}$. The final two-class prediction (bona fide or spoofed) is then obtained using projection and output layers.
\section{Experimental setup }
\label{section:exp_setup}
Described in this section are the database, metrics and baseline systems used for our experiments, together with specific implementation details of the RawGAT-ST model.

\subsection{Database and evaluation metric}
Experiments were performed on the ASVspoof 2019 logical access (LA) database~\cite{wang2020asvspoof,todisco2019asvspoof}. 
It has three independent partitions: train; development; evaluation. 
Spoofed speech is generated using a set of different speech synthesis and voice conversion algorithms~\cite{wang2020asvspoof}. 
The training and development partitions were created with a set of 6 different attacks (A01-A06), whereas the evaluation set was created with a set of 13 attacks (A07-A19). 
We used the minimum normalised tandem detection cost function (t-DCF)~\cite{kinnunen2018t,kinnunen-tDCF-TASLP} as a primary metric but also report results in terms of the pooled equal error rate (EER)~\cite{bosaris}.

\subsection{Baseline}
The baseline is an end-to-end RawNet2 system~\cite{tak2021rawnet}.  It is among the best-performing, reproducible solutions. 
The first sinc layer is initialised with a bank of 128 mel-scaled filters. 
Each filter has an impulse response of 129 samples (~8~ms duration) which is convolved with the raw waveform. The latter are truncated or concatenated to give segments of approximately {$4$}~seconds duration (64,600 samples). 
It is followed by a residual network and a gated recurrent unit (GRU) to predict whether the input audio is bona fide or spoofed speech. 
Full details of the baseline system are available in~\cite{tak2021rawnet}.

Our temporal and spectral GAT systems introduced in~\cite{tak2021graph} are not used as baselines in this work since they are not E2E systems; unlike the spectro-temporal GAT system introduced in this paper, they operate upon hand-crafted features. 
Nonetheless, we do report results for E2E temporal and spectral GAT solutions which are based upon the appropriate sub-blocks illustrated in Fig.~\ref{fig:GAT pipeline}. 
Use of these temporal and spectral GAT variants is necessary in order that differences in performance can be attributed solely to use of the proposed spectro-temporal GAT model and not also from differences in the first network layer. 
Details are described in Section~\ref{sec:Ablation Study}.

\subsection{RawGAT-ST implementation} 
\label{sec:implmentation}

In contrast to the baseline RawNet2 system and in order to reduce computational complexity, we reduced the number of filters in the first layer to $70$.
To improve generalisation, we added channel masking in similar fashion to the frequency masking~\cite{park2019specaugment,chen2020generalization,wang2021specaugment++} to mask (set to zero) the output of a random selection of contiguous sinc channels during training. 
The same channel mask is applied to all training data within each mini-batch. The number of masked channels is chosen from a uniform distribution between $0$ and $F_{\textrm{mask}}$, where $F_{\textrm{mask}}=14$ is the maximum number of masked channels selected based on minimum validation loss. 
While attention helps the model to focus on the most discriminative spectral sub-bands and temporal segments, channel masking improves generalization by ensuring that information at all sub-bands and segments is used at least to some extent.
In contrast to usual practice, we also use fewer filters ($32$ and $64$) in the first and second residual blocks to further protect generalisation to unseen attacks~\cite{parasu2020investigating}. 
Graph pooling is applied with empirically selected pooling ratios of $k$ = 0.64, 0.81, and 0.64 for spectral, temporal and spectro-temporal attention blocks respectively. 

The complete architecture is trained using the ASVspoof 2019 LA training partition to minimise a weighted cross entropy (WCE) loss function, where the ratio of weights assigned to bonafide and spoofed trials are 9:1 to manage the data imbalance in the training set. 
We used the standard Adam optimiser~\cite{kingma2014adam} with a mini-batch size of $10$ and a fixed learning rate of $0.0001$ and train for $300$ epochs. The feature extractor and back-end classifier are jointly optimised using back-propagation~\cite{Goodfellow_2016}. 
The best model was selected based on the minimum validation loss (loss on development set).
The spectro-temporal GAT model has only 0.44M parameters and is comparatively light weight compared to the baseline as well as other state of the art systems. 
All experiments were performed on a single GeForce RTX 3090 GPU and reproducible with the same random seed and GPU environment using an open source implementation~\footnote{https://github.com/eurecom-asp/RawGAT-ST-antispoofing}.

\section{Experimental Results}
\label{sec:experimental results}
In this section, we present our results, an ablation study and a comparison of our results to those of existing, state-of-the-art single systems. 

\begin{table*}[!ht]
\def\arraystretch{1.1}
	\small
	\centering
	
	\caption{Results for the ASVspoof 2019 logical access (LA) database in terms of min t-DCF for each attack in the evaluation  partition (A07-A19).  Pooled min~t-DCF (P1) and pooled EER (P2) are shown in the last two columns.  Results shown for the  end-to-end RawNet2 baseline system and the three RawGAT-ST based systems introduced in this paper.}
	
    \vspace{0.2cm}
	\setlength\tabcolsep{4.3pt}
	\begin{tabular}{c || *{13}{c}| cc} 
		\hline
		
		System &A07&A08&A09&A10&A11&A12&A13&A14&A15&A16&A17&A18 &A19& P1&P2 	\\ 
		\hline\hline
        RawNet2-baseline~\cite{tak2021rawnet,tak2021graph}&.098&.179&.073&.089&.042&.088&.020&.013&.073&.046&.240&.629&.058&0.1547&5.54\\
		\hline
		RawGAT-ST-add&.011&.030&.004&.016&.009&.019&.007&.007&.014&.027&.053&.104&.023&0.0373&1.15\\
		\hline
		RawGAT-ST-concat&.021&.027&.003&.027&.008&.029&.015&.008&.022&.029&.046&.120&.022&0.0388&1.23\\
		\hline
		RawGAT-ST-mul&.010&.016&.002&.012&.010&.010&.010&.009&.009&.023&.055&.080&.024&\textbf{0.0335}&\textbf{1.06}\\
		\hline
	\end{tabular}
	\label{Tab: results RawGAt eval}
\end{table*}

\vspace{-0.2cm}
\subsection{Results}

Results
are illustrated in 
Table~\ref{Tab: results RawGAt eval}, where the last two columns P1 and P2 indicate pooled min t-DCF and pooled EER results for the baseline system (RawNet2) and three different configurations of the new RawGAT-ST system. 
The variants involve the use of different spectro-temporal fusion strategies.
%
%
Whereas all RawGAT-ST systems outperform the baseline by a substantial margin, the best result is obtained using the RawGAT-ST-mul system for which the t-DCF is 0.0335 (cf.\ 0.1547 for the baseline) and the EER is 1.06\% (5.54\%).  
%
%

These results show that 
all RawGAT-ST systems are effective in exploiting spectro-temporal attention and to   
model the relationships between different spectro-temporal estimates, thereby improving the discrimination between spoofed and bona fide inputs. We now seek to demonstrate the merit of early fusion at the model level, rather than late fusion at the score level. This is done by way of ablation experiments.

\subsection{Ablation study}
\label{sec:Ablation Study}
Only through ablation experiments can we properly demonstrate the merit of the RawGAT-ST approach; we cannot use results from our previous work~\cite{tak2021graph} since they were generated using hand-crafted features.
Through ablation, we essentially remove one of the blocks or operations in the full RawGAT-ST architecture illustrated in Fig.~\ref{fig:GAT pipeline}.  The remaining blocks and operations are then used as before.  Results are illustrated in Table~\ref{Tab:ablation results}.  The middle row highlighted in boldface is the RawGAT-ST-mul result that is also illustrated in Table~\ref{Tab: results RawGAt eval}.


Ablation of the spectral GAT attention block (top left in Fig.~\ref{fig:GAT pipeline}) leaves the system being capable of exploiting only temporal attention.  Without spectral attention (first row of Table~\ref{Tab:ablation results}), performance degrades by 34\% relative to the full system (0.0514 cf.\ 0.0335).  
The degradation without temporal attention (0.0385) is less severe (13\% relative), indicating the greater importance of spectral attention versus temporal attention, even if both are beneficial.
Last, we demonstrate the benefit of graph pooling by ablating the pooling layers in all three blocks.
The relative degradation in performance of 58\% (0.0788 cf.\ 0.0335) is even more substantial and shows the benefit of using graph pooling to 
concentrate on the most informative node features.  

\begin{table}[h]
	\centering
	\small

	\caption{Results for ablation studies }
	\vspace{0.3cm}
	\renewcommand{\arraystretch}{1.1}
 \setlength\tabcolsep{2.75pt}
	\begin{tabular}{ *{4}{c}}
		\hline
		 System & min-tDCF& EER	\\ 
	\hline\hline
	w/o spectral attention&0.0514&1.87\\
	\hline
	w/o temporal attention&0.0385&1.13\\
	\hline
	\textbf{w/ spectro-temporal attention} & \textbf{0.0335} & \textbf{1.06}\\
	\hline
	w/o graph pooling&0.0788&2.47\\

		 \hline
	\end{tabular}
	\label{Tab:ablation results}
	\vspace{-0.15cm}
\end{table}

\subsection{Performance comparison}

Illustrated in Table~\ref{Tab:comparsion results} is a comparison of performance for the proposed RawGAT-ST system and competing single systems reported in the literature~\cite{nautsch2021asvspoof}. 
The comparison shows that our system which uses GATs with self attention outperforms 
alternative attention approaches such as:
Convolutional Block Attention Module (CBAM); Squeeze-and-Excitation (SE); Dual attention module with pooling and convolution operations. 
Furthermore, to the best of our knowledge, our approach is the best single model system reported in the literature to date.
Our system is also one of only two that operates on the raw signal.  With only 0.44M parameters, the proposed RawGAT-ST system is also among the least complex; only the Res-TSSDNet (0.35M)~\cite{hua2021towards} and LCNN-LSTM-sum (0.27M)~~\cite{wang2021comparative} have fewer parameters, while our system performs better.
%
%
Despite the simplicity, the RawGAT-ST system outperforms our previous best temporal GAT system with late score fusion~\cite{tak2021graph} (row 12 in Table~\ref{Tab:comparsion results}) by
63\% and 77\% relative in terms of min t-DCF and EER respectively.
This result also points towards the benefit of operating directly upon the raw signal in fully end-to-end fashion.

\begin{table}[!t]
	\centering
	\small

	\caption{Performance for the ASVspoof 2019 evaluation partition in terms of pooled min t-DCF and pooled EER for different state-of-the-art single systems and the best performing RawGAT-ST-mul system (boldface).  Only the top two RawGAT-ST and Res-TSSDNet systems operate directly upon raw inputs; all others use some form of hand-crafted representations.}
	\vspace{0.3cm}
 \setlength\tabcolsep{3pt}
  \renewcommand{\arraystretch}{1.1}
	\begin{tabular}{ *{4}{c}}
		\hline
		 System & front-end & min-tDCF& EER	\\ 
	\hline	\hline
	 \textbf{RawGAT-ST (mul)}&\textbf{Raw-audio}&\textbf{0.0335}&\textbf{1.06}\\
	 
	 \hline	
	 Res-TSSDNet~\cite{hua2021towards}&Raw-audio&0.0481&1.64\\ 
	 \hline	
	 MCG-Res2Net50+CE~\cite{li2021channelwise}&CQT&0.0520&1.78\\
	  \hline	
	 ResNet18-LMCL-FM~\cite{chen2020generalization}&LFB&0.0520&1.81\\
	 \hline
	LCNN-LSTM-sum~\cite{wang2021comparative}&LFCC&0.0524&1.92\\  
	 \hline	
	 Capsule network~\cite{luo2021capsule}&LFCC&0.0538&1.97\\  
	\hline
	
	Resnet18-OC-softmax~\cite{zhang2021one}&LFCC&0.0590&2.19\\
	\hline
	MLCG-Res2Net50+CE~\cite{li2021channelwise}&CQT&0.0690&2.15\\
	\hline
	SE-Res2Net50~\cite{li2021replay}&CQT&0.0743&2.50\\
	\hline
	LCNN-Dual attention~\cite{ma2021improved}&LFCC&0.0777&2.76\\
	\hline
	Resnet18-AM-Softmax~\cite{zhang2021one}&LFCC&0.0820&3.26\\
	\hline
	ResNet18-GAT-T~\cite{tak2021graph}&LFB&0.0894&4.71\\
		 \hline
	GMM~\cite{IS2020LFCC} & LFCC&0.0904&3.50\\
		 \hline
		 ResNet18-GAT-S~\cite{tak2021graph}&LFB&0.0914&4.48\\
		 \hline
	PC-DARTS~\cite{ge2021}&LFCC&0.0914&4.96\\
		 \hline
		 Siamese CNN~\cite{lei2020siamese}&LFCC& 0.0930&3.79\\
	\hline
LCNN-4CBAM~\cite{ma2021improved}&LFCC&0.0939&3.67\\
		 \hline

	LCNN+CE~\cite{das2021data}&DASC-CQT&0.0940&3.13\\
	
	\hline
	 LCNN~\cite{lavrentyeva2019stc}&LFCC&0.1000&5.06\\
		 \hline
	LCNN+CE~\cite{wu2020light}&CQT&0.1020&4.07\\
		 \hline

		  ResNet~\cite{yang2021modified}&CQT-MMPS&0.1190&3.72\\
		 \hline
	\end{tabular}
	\label{Tab:comparsion results}
\end{table}

\section{Conclusions}
\label{sec:conclusion}

Inspired from previous findings which show that different attention mechanisms work more or less well for different forms of spoofing attack, we report in this paper a novel graph neural network approach to anti-spoofing and speech deepfake detection based on model-level spectro-temporal attention. 

The RawGAT-ST solution utilises a self attention mechanism to learn the relationships between different spectro-temporal estimates and the most discriminative nodes within the resulting graph.  
The proposed solution also operates directly upon the raw waveform, without the use of hand-crafted features, and is among the least complex of all solutions reported thus far.
Experimental results for the standard ASVspoof 2019 logical access database show the success of our approach and are the best reported to date by a substantial margin.
Results for an ablation study show that spectral attention is more important than temporal attention, but that both are beneficial, and that graph pooling improves performance substantially.  Last, a comparison of different systems, working either on the raw signal or hand-crafted inputs show the benefit of a fully end-to-end approach.   


We foresee a number of directions with potential to improve performance further. These ideas include the use of learnable pooling ratios to optimise the number of node features retained after graph pooling and the use of a learnable weighted selective fusion module to dynamically select the most discriminative features from spectral and temporal attention networks. Working with the latest ASVspoof 2021 challenge database~\cite{ASV2021challenge}, these ideas are the focus of our ongoing work.


\section{Acknowledgements}
This work is supported by the VoicePersonae project funded by the French Agence Nationale de la Recherche (ANR) and the Japan Science and Technology Agency (JST).

\vfill

\clearpage
\balance
\bibliographystyle{IEEEbib}
\bibliography{ASVspoof2021_BibEntries,refs,GAT_IS_21,GAT_jee_ref}


\end{document}